\documentclass[aps,pre,amsmath,amssymb,superscriptaddress,floatfix,showpacs,twocolumn]{revtex4}

\usepackage{amsmath}
\usepackage{amssymb}
\usepackage{graphicx}
\usepackage{dcolumn}
\usepackage{bm}

\input epsf.sty

\makeatletter
\def\captionof#1#2{{\def\@captype{#1}#2}}
\makeatother

\def\AA{{\cal A}}

\def\JJ{{\cal J}}

\def\PP{{\cal P}}

\def\Cg{{\bf C}}
\def\Dg{{\bf D}}
\def\Eg{{\bf E}}
\def\Pg{{\bf P}}

\def\Ks#1#2{K^{*#1}_{\hphantom{*}#2}}
\def\As#1#2{A^{*#1}_{\hphantom{*}#2}}

\def\moyenne#1{\langle #1\rangle}

\def\normK{\left|\left| K^*\right|\right|}
\def\NK{\left|\left| K^*\right|\right|_1}

\begin{document}

\title{A measure of the violation of the 'detailed balance' criterion:\\
a possible definition of a 'distance' from equilibrium}
\author{T. Platini}
\affiliation{Department of Physics and Virginia Bioinformatics Institute, Virginia Polytechnic Institute and State University, Blacksburg, Virginia 24061-0435, USA}


\begin{abstract}
Motivated by the classification of non-equilibrium steady states suggested by R.K.P. Zia and B. Schmittmann in J. Stat. Mech. P07012 (2007), we propose to measure the violation of the 'detailed balance' criterion by the $p$-norm ($\normK_p$) of the matrix formed by the probability currents. Its asymptotic analysis, for the totally asymmetric exclusion process, motivates the definition of a 'distance' from equilibrium $K^*$ obtained for $p=1$. In addition, we show that the latter quantity and the average activity $\moyenne{\AA^*}$ are both related to the probability distribution of the entropy production. Finally, considering the open-ASEP and open-ZRP, we show that the current of particles gives an exact measure of the violation of 'detailed balance'. 
\end{abstract}

\date{\today}
\maketitle
\section{Introduction}
In nature, typically every system is governed by well-known 'deterministic' physical laws. However the microscopic details are usually unreachable and a full description of the system is impossible. The enormous number of degrees of freedom and the apparent 'chaotic' motion of the microscopic elements leads to rather complicated phenomenon. To face such situations, the physics of statistical mechanics has been built over the last two centuries. Our best approximation is to assume that the interactions between the microscopic elements occur according to some probabilistic rules. It follows that the natural reformulation of many-body problems takes the form of stochastic processes \cite{Kampen}. In addition, almost all systems are in interaction with the environment. It is only in particular limits (for defined time and spatial scales) that closed systems emerge. Generally, currents of particles, heat or magnetization are induced by the environment and testify to the non-equilibrium activity. If the presence of flux is a signature of a non-equilibrium steady state (NESS), the absence of macroscopic currents does not necessarily imply that the system is at equilibrium. By definition equilibrium is characterized, on a microscopic level, by the well-known 'detailed balance' criterion. The latter 'balance' is breaking for any non-equilibrium state and imposes non-vanishing currents of probability flowing between connected states.\\
If equilibrium systems are well described by the theory of ensembles, no global formalism exists for systems out of equilibrium. Recently, general relations for driven systems have attracted a lot of interest. One should mention the Kawasaki relation \cite{Kawasaki-1967}, the Jarzynski and Crooks relations \cite{Jarzynski-1997a,Jarzynski-1997b,Crooks-1998, Crooks-2000} and 'fluctuation relations' that include the fluctuation theorems of Evans-Searles \cite{Evans-Searles-1994,Evans-Searles-1996} and Gallavotti-Cohen \cite{Gallavotti-1995a, Gallavotti-1995b}. While analyzing NESS case-by-case, research works are traditionally focused on the probability distribution $\{\Pg^*\}$ of the micro-states. This approach is usually relevant since almost all observables can be extracted from there. However, the analysis of currents or average production of entropy requires the knowledge of the transition rates which complete the characterization of the steady state. Recently, Zia and Schmittmann \cite{RoyceA,RoyceB} suggested a general classification of NESS, where a complete description of the system is given by the distribution of the probabilities and probability currents $\left\{P^*,K^*\right\}$. This classification allows the identification of the transformations of the transition rates that leave the steady state invariant. Along these lines, a set of invariant quantities has been derived for a class of steady states driven by the boundaries dynamics \cite{Baule-2008}. From the probability currents a definition of the 'Euclidean distance' from equilibrium has been proposed in \cite{RoyceA}. In this paper we suggest to measure the violation of 'detailed balance' and define the 'distance' from equilibrium by the $p$-norm ($||K^*||_p$) of the matrix formed by the probability currents. We give, for the periodic totally asymmetric simple exclusion process (TASEP), the exact expression of $||K^*||_p$. We show, in the thermodynamic limit, that the $p$-norm vanishes for any $p\ne1$. This result motivates the definition of the 'distance' from equilibrium by $\NK$ that we prove to be extensive for the periodic-TASEP, open-ASEP and open-ZRP. One should mention that the $1$-norm was first defined in \cite{Dorosz-2009} and used to measure the violation of the 'detailed balanced' criterion for different reaction-diffusion models.\\

We start our paper with a presentation of the master equation which governs the time evolution of Markovian stochastic systems. Based on the analysis of the asymptotic behavior of the $p$-norm $||K^*||_p$, for the periodic-TASEP, we define the 'distance' from equilibrium.  In section III, we show that the average activity $\moyenne{\AA^*}$ and the 'distance' from equilibrium $K^*$ are both related to the production of entropy. In section IV, for the open-ASEP and open-ZRP, we show that the current of particles gives an exact measure of the violation of 'detailed balance'. Finally our results are summarized in section V.
\section{General Framework}
When describing Markovian stochastic many body systems, the master equation is the most general equation that one has at his disposal. Its formalism is used to describe in and out of equilibrium systems, such as chemical reactions, enzyme kinetics, biological populations etc... . In this section, we give a presentation of the master equation and introduce the matrix formed by the probability currents. For the periodic-TASEP we discuss, in the thermodynamic limit, the asymptotic scaling of the $p$-norm $\normK_p$ that motivates the definition of the 'distance' from equilibrium by $\NK$.
\subsection{Master Equation}
Considering the continuous dynamics of a many-body system, we note $X$ a particular state of the configuration space. The probability $P_X(t)$, to find the system in the state $X$ at time $t$, is governed by the master equation 
\begin{eqnarray}\label{MasterEq}
\partial_t P_X(t)=\sum_{Y\ne X}K^Y_X(t),
\end{eqnarray}
with the net probability current
\begin{eqnarray}
K^Y_X(t)=w^Y_XP_Y(t)-w^X_YP_X(t),
\end{eqnarray}
where $w^X_Y$ is the rate of the transition from the configuration $X$ to $Y$. We assume the dynamics to be ergodic which imposes that the system has a unique stationary state independent of the initial condition. The master equation (\ref{MasterEq}) expresses the time derivative of the probability as the balance of the currents flowing in and out the state $X$. Defining $J_+(X)$ and $J_-(X)$ as
\begin{eqnarray}
J_+(X)=\sum_{Y\ne X}w^Y_XP_Y(t),\ \ J_-(X)=\sum_{Y\ne X}w^X_YP_X(t),
\end{eqnarray}
the equation  (\ref{MasterEq}) simply states the conservation of probability through the expression $\partial_t P_X(t)=J_+(X)-J_-(X)$. In the steady state, the current appears to be 'globally balanced' on each state so that $J_+(X)=J_-(X)$. By definition, all equilibrium distributions verify the 'detailed balance' criterion
\begin{eqnarray}
\Ks{Y}{X}=w^Y_XP^*_Y-w^X_YP^*_X=0.
\end{eqnarray}
However, for NESS this criterion is locally violated and the probability current on each bond $\Ks{X}{Y}$ is not zero. In the network, where nodes are states of the system and links are possible transitions between states, the violation of 'detailed balance' imposes the existence of loops of current. In such a situation the system presents generally (but not always) macroscopic currents.
\subsection{The $p$-norm as a 'distance' from equilibrium}
In the stationary state, a natural definition of the 'distance' from equilibrium is given by the norm of the matrix defined by the elements $\Ks{Y}{X}$. For any systems evolving in a finite configuration space, we define the $p$-norm by
\begin{eqnarray}
\normK_p=\left(\frac{1}{2}\sum_{X,Y}\left|\Ks{Y}{X}\right|^p\right)^{1/p}, & p\in{\mathbb N}^*
\end{eqnarray}
where the factor $1/2$ has been introduced for convenience. This quantity measures the violation of 'detailed balance' and vanishes at equilibrium only. In references \cite{RoyceA,RoyceB} the authors suggested to define an 'Euclidean distance' from equilibrium by $\normK_2$. In addition, they showed that each quadratic element $|\Ks{X}{Y}|^2$ appears in the expression of the entropy production. \\

Choosing the periodic-TASEP as an example, we obtain an exact expression of $\normK_p$ as a function of the average current of particles $\moyenne{j^*}$. Let us remind the reader that the latter model is defined on a one-dimensional periodic lattice, where particles are strictly jumping forward to the next nearest neighboring site. The transitions are only possible if the target site is empty and occur with a rate equal to one. Our work is facilitated by the fact that the stationary probabilities are equiprobable and that the dynamical process is totally asymmetric ($w^Y_X=0$ if $w^X_Y=1$). Our calculations give for the $p$-norm
\begin{eqnarray}
\normK_p=\left(\frac{L\moyenne{j^*}}{\Omega^{p-1}}\right)^{1/p},
\end{eqnarray}
where $\Omega$ is the dimension of the configuration space and $L$ is the size of the system. As a consequence, while taking the thermodynamic limit and keeping the density of particles constant, we can show that $\normK_p$ vanishes for all $p\ne1$. Writing $\rho$ the average density of particles, the asymptotic analysis ($L>>1$) gives us
\begin{eqnarray}
\normK_p\propto \sqrt{L^{1+1/p}}\times e^{-sL(1-1/p)},
\end{eqnarray}
where 
\begin{eqnarray}
s=-\rho\ln\rho-(1-\rho)\ln(1-\rho).
\end{eqnarray}
In this particular case, as in many others, the total current of particles $\moyenne{\JJ^*}$ appears as the simplest observable that characterized the non-equilibrium behavior of the system. This quantity increases with the system size while its density $\moyenne{j^*}=\moyenne{\JJ^*}/L$ remain constant. In our example, and with a natural understanding of the 'distance' from equilibrium, the limit $\normK_p\rightarrow0$ as $L\rightarrow\infty$ is clearly inadequate. The only 'satisfying' definition consists of defining the 'distance' from equilibrium using the $1$-norm. For simplicity, we will note hereafter $K^*=\NK$. Exactly derived for the periodic-TASEP, open-ASEP and open-ZRP, $K^*$ appears to be extensive. Its relations to the activity and entropy production are presented in the following sections. 
\section{Activity and entropy production}
If $\Ks{X}{Y}$ is well understood as the probability current, usually little is said about the role of the symmetric part $\As{X}{Y}=w^X_Y P^*_X+w^Y_X P^*_Y$ and its properties. Introduced in \cite{Maes-2006,Lecomte-2007}, only recently has this quantity been analyzed, successively called traffic \cite{Maes-2008a,Maes-2008b,Maes-2008c}, dynamical activity \cite{Appert-Rolland-2008,Bodineau-2008,Garrahan-2009,Gorissen-2009} or frenesy. Its central role in the non-equilibrium linear response theory and out-of-equilibrium dynamical fluctuation theory has been shown in \cite{Baiesi-2009,Baiesi-20010}. Using the inequality $\left|\Ks{X}{Y}\right|\le \As{X}{Y}$, we trivially obtain the inequality
\begin{eqnarray}\label{Inequality}
K^*\le\moyenne{\AA^*},
\end{eqnarray}
with $\moyenne{\AA^*}=\frac{1}{2}\sum_{X,Y\ne X}A^{*X}_{\hphantom{*}Y}$ which can be expressed as the average of the observable $\AA(X)$ defined as
\begin{eqnarray}
\AA(X)=\sum_{Y\ne X} w^X_Y.
\end{eqnarray}
The latter quantity corresponds to the escape rate and tells us how 'nervous' is the system in a given state. If one thinks about a system which presents, in its stationary state, a macroscopic current $\moyenne{{\cal J}^*}$, the equation ($\ref{Inequality}$) should be understood as a generalization of the inequality $\moyenne{{\cal J}^*}=\moyenne{{\cal J}_+^*}+\moyenne{{\cal J}_-^*}\le|\moyenne{{\cal J}_+^*}|+|\moyenne{{\cal J}_-^*}|$, where $\moyenne{{\cal J}_+^*}$ and $\moyenne{{\cal J}_-^*}$ are the averages of the currents flowing in opposite directions.\\ 
As already noticed in \cite{RoyceA, RoyceB}, the activity and the current of probability appear explicitly in the expression of total entropy production. Defining by $\moyenne{\delta S}_\tau$ the average production of entropy measured over a time interval $\tau$ ($\tau\ll1$), the rate of creation $\sigma=\moyenne{\delta S}_\tau/\tau$ is given by
\begin{eqnarray}
\sigma=\sum_{X,Y}P_X^*w^X_Y\ln\left(\frac{P^*_Xw^X_Y}{P^*_Yw^Y_X}\right),
\end{eqnarray}
which can be explicitly written as
\begin{eqnarray}
\sigma =\frac{1}{2}\sum_{X,Y}\Ks{X}{Y}\ln\left(\frac{\As{X}{Y}+\Ks{X}{Y}}{\As{X}{Y}-\Ks{X}{Y}}\right).
\end{eqnarray}
To pursue further we define the probability $\PP_\tau(a,k)$ to measure, in the stationary state, over a time interval $\tau$ ($\tau\ll1$), an activity $a$ and a current of probability $k$. By definition, $\PP_\tau(a,k)$ is
\begin{eqnarray}\label{EQPROBA}
\PP_\tau(a,k)=\sum_{X,Y}P^*_Xw^X_Y\tau\hphantom{.}\delta\left(\As{X}{Y}-a\right)\delta\left(\Ks{X}{Y}-k\right).\nonumber\\
\end{eqnarray}
Using the relations $\As{X}{Y}=\As{Y}{X}$ and $\Ks{X}{Y}=-\Ks{Y}{X}$, we prove easily 
\begin{eqnarray}
(a+k)\PP_\tau(a,-k)=(a-k)\PP_\tau(a,k),
\end{eqnarray}
which can be written as
\begin{eqnarray}\label{EQProba}
\frac{\PP_\tau(a,k)}{\PP_\tau(a,-k)}=e^{\delta S},\ \text{with} \ \delta S=\ln\left(\frac{a+k}{a-k}\right).
\end{eqnarray}
The probability $\PP_\tau(\delta S)$ of a creation of entropy $\delta S$, over a time interval $\tau$ ($\tau\ll1$), is given by
\begin{eqnarray}
\PP_\tau(\delta S)=\int_0^{+\infty}da\int_{-\infty}^{+\infty}dk\hphantom{.}\PP_\tau(a,k)\hphantom{.}\delta\left[\delta S-\ln\left(\frac{a+k}{a-k}\right)\right],\nonumber\\
\end{eqnarray}
and such that equation ($\ref{EQProba}$) leads to the well known ratio $P_\tau(\delta S)/P_\tau(-\delta S)=e^{\delta S}$.\\

Particularly interesting results emerge from the analysis of the probabilities 
\begin{eqnarray}
\PP_\tau(\delta S<0)&=&\int_{-\infty}^{0^-} d(\delta S) \hphantom{.}\PP_\tau(\delta S)\\
\PP_\tau(\delta S>0)&=&\int_{0+}^{+\infty} d(\delta S)\hphantom{.} \PP_\tau(\delta S),
\end{eqnarray}
respectively, probability of a positive and negative creation of entropy. From equation ($\ref{EQPROBA}$) one easily show 
\begin{eqnarray}\label{SarahLW}
\PP_\tau(\delta S\gtrless0)&=&\sum_{X,Y}P^*_Xw^X_Y\tau\hphantom{.}\Theta\left(\pm \Ks{X}{Y}\right),
\end{eqnarray}
where $\Theta(x)$ is the Heaviside function defined by $\Theta(x)=1$ for $x>0$ and $\Theta(x)=0$ otherwise. Expressed as a function of $\As{X}{Y}$ and $\Ks{X}{Y}$, equation ($\ref{SarahLW}$) takes the form
\begin{eqnarray}
\PP_\tau(\delta S \gtrless0)&=&\frac{\tau}{4}\sum_{X,Y}\As{X}{Y}\left[\Theta\left(\pm\Ks{X}{Y}\right)+\Theta\left(\mp\Ks{X}{Y}\right)\right]\notag \\
&+&\frac{\tau}{4}\sum_{X,Y}\Ks{X}{Y}\left[\Theta\left(\pm\Ks{X}{Y}\right)-\Theta\left(\mp\Ks{X}{Y}\right)\right],\notag \\
\end{eqnarray}
Restraining ourself to systems for which 'detailed balance' is violated for every transitions such that $\Ks{X}{Y}\ne0$  $\forall X\ne Y$, one finally obtains
\begin{eqnarray}\label{sigma+}
\PP_\tau(\delta S \gtrless0)&=&\frac{\moyenne{\AA^*}\pm K^*}{2}\tau.
\end{eqnarray}
As a consequence, the probability $\PP_\tau(\delta S=0)$ is solely expressed as a function of the activity through the equality
\begin{eqnarray}
\PP_\tau(\delta S=0)=1-\moyenne{\AA^*}\tau.
\end{eqnarray}
In other words, the probability of a non-zero production of entropy is given by the activity produced over the time interval $\tau$. This is expressed under the equality
\begin{eqnarray}
\moyenne{A^*}=\frac{\PP_{\tau}(\delta S\ne0)}{\tau}.
\end{eqnarray} 
Along the same line, the distance from equilibrium is measured by the difference
\begin{eqnarray}\label{LittleBritish}
K^*=\frac{\PP_\tau(\delta S>0)-\PP_\tau(\delta S<0)}{\tau}.
\end{eqnarray}
The latter equality can be explicitly expressed as the average of the function sign
\begin{eqnarray}
K^*=\frac{1}{\tau} \int_{-\infty}^{+\infty}d(\delta S)\hphantom{.} \PP_\tau(\delta S) sign(\delta S),
\end{eqnarray}
where $sign(x)$ is defined by
	\begin{eqnarray}
	sign(x)=\left\{
	\begin{array}{cc}
	-1&x<0\\
	0& x=0\\
	+1& x>0.
	\end{array}\right.
	\end{eqnarray}
Since $K^*>0$ we trivially obtain $\moyenne{sign(\delta S)}>0$, which simply states the inequality $\PP_{\tau}(\delta S >0)>\PP_{\tau}(\delta S <0)$.

\section{Driven systems}
If the calculation of the 'distance' $K^*$ can appear challenging, exact results are reachable for simple driven systems. In this section, we give the keys of the derivation which leads to the exact expression of $K^*$ for the open-ASEP and ZRP. These two systems have been the favorite toy models of the scientific community. They are some of the rare exactly solvable models, used as guides in the zoology of exotic behaviors found in the field of non-equilibrium many body systems. For these two cases, it is the particular structure of the stationary probabilities that allows an exact calculation of $K^*$.\\

As a preliminary result, from the inequality (\ref{Inequality}), one show that $K^*$ has to satisfy
\begin{eqnarray}\label{EqOPEN}
K^*\le|\moyenne{\JJ_+^*}|+|\moyenne{\JJ_-^*}|,
\end{eqnarray}
where $\moyenne{\JJ_{\pm}^*}$ are the average currents defined by
\begin{eqnarray}
\moyenne{\JJ_\pm^*}=\sum_{X,Y}P^*_Xw^X_Y \delta\left(X {\overset{\pm}{\rightarrow}}Y\right),
\end{eqnarray}
where $\delta\left(X {\overset{\pm}{\rightarrow}}Y\right)=1$ if the transition between the states $X$ and $Y$ is induced by the jump forward/backward of a particle. The total current of particles is given by the sum $\moyenne{\JJ^*}=\moyenne{\JJ_+^*}+\moyenne{\JJ_-^*}$. For such systems that do not present other reactions than the biased diffusion, the current of particles in the stationary state is directly responsible of the production of entropy. Therefore a vanishing current would imposes $\moyenne{\delta S}=0$ so that we expect $K^*=K^*(\moyenne{\JJ^*})$ with $K^*(0)=0$.
\subsection{Open-ASEP}
The ASEP is the simplest one-dimensional driven model that includes the biased diffusion of hard-core particles. Traditionally, the left and right diffusion rates are respectively denoted by $q$ and $p$. Without loss of generality, we are considering $p>q$ which imposes $\moyenne{j^*}>0$. The exclusion process imposes that the jump of a particle is only possible if the target site is empty. The system is driven at its boundaries by the interaction with two reservoirs of particles. On the left extremity, particles can be injected or removed with rates $\alpha$, respectively $\gamma$. Identically, on the right boundary, the rates of injection and ejection are written $\delta$ and $\beta$. The open-TASEP is recovered for $p=1$, $q=0$ and $\gamma=\delta=0$. It is well known that the exact expression of the stationary state can be expressed as the product of non-commuting matrices \cite{Derrida-1992,Derrida-1993}. For a state $X$ defined by the set of occupation numbers $\{\eta_1(X),\hdots,\eta_L(X)\}$, the stationary probability  can be written as $P^*_X=f_L(X)/Z_L$ with
\begin{eqnarray}
f_L(X)=\langle W| \prod_{n=1}^L\left[\eta_n(X)\Dg+\left\{1-\eta_n(X)\right\}\Eg\right]|V\rangle,
\end{eqnarray}
and $Z_L=\sum_{X}f_L(X)$. The matrices $\Dg$ and $\Eg$ satisfy
\begin{eqnarray}
\label{EqCom1}
p\Dg\Eg-q\Eg\Dg=\Dg+\Eg=\Cg,
\end{eqnarray}
and act on the vectors $\langle W| $ and $|V\rangle$ as 
\begin{eqnarray}
\label{EqCom2}
& &(\beta\Dg-\delta\Eg)|V\rangle=|V\rangle\\
\label{EqCom3}
& &\langle W|(\alpha\Eg-\gamma\Dg)=\langle W|.
\end{eqnarray}
It is important to note that the 'distance' from equilibrium can be decomposed over the bulk ($B^*$) and boundaries contributions. We note $K^*=L^*+B^*+R^*$, where $L^*$ and $R^*$ are respectively the left and right boundary terms. Since for each transition that is induced by the jump forward of a particle ($w^X_Y=p$) the reversed transition rate is $w^Y_X=q$, the bulk contribution takes the form
\begin{eqnarray}
B^*=\frac{1}{Z_L}\sum_{X,Y}|pf(X)-qf(Y)|\delta\left(X {\overset{+}{\rightarrow}}Y\right).
\label{EqBulk}
\end{eqnarray}
Considering a particle initially localized on the site $n$, the function $f(X)$ and $f(Y)$ can be written as 
\begin{eqnarray}
f_L(X)&=&\langle W| \Phi_{n}(X) \Dg\Eg \Psi_{n}(X)|V\rangle,\\
f_L(Y)&=&\langle W| \Phi_{n}(X) \Eg\Dg \Psi_{n}(X)|V\rangle,
\end{eqnarray}
where $\Phi_{n}(X)=\prod_{i=1}^{n-1}\left[\eta_i(X)\Dg+\left\{1-\eta_i(X)\right\}\Eg\right]$ and $\Psi_{n}(X)=\prod_{i=n+2}^{L}\left[\eta_i(X)\Dg+\left\{1-\eta_i(X)\right\}\Eg\right]$. Using the commutation rule (\ref{EqCom1}) the contribution $B^*$ is given by
\begin{eqnarray}
B^*&=&\frac{1}{Z_L}\sum_{n}\sum_{X}\langle W| \Phi_{n}(X) \Cg \Psi_{n}(X) |V\rangle.
\end{eqnarray}
Since the bracket sandwich expresses a probability, each term of the sum is positive and the absolute value can be omitted. This last result leads to $B^*=(L-1)Z_{L-1}/Z_{L}$, where the ratio $Z_{L-1}/Z_{L}$ corresponds to the average current of particles $\moyenne{j^*}$. For details on the algebraic method see \cite{Derrida-1992,Derrida-1993}. Along the same line, one has for each boundary,
\begin{eqnarray}
L^*&=&\frac{1}{Z_L}\sum_{X,Y}|\alpha f(X)-\gamma f(Y)|\delta\left(X {\overset{\alpha}{\rightarrow}}Y\right),\\
\label{EqA}
R^*&=&\frac{1}{Z_L}\sum_{X,Y}|\beta f(X)-\delta f(Y)|\delta\left(X {\overset{\beta}{\rightarrow}}Y\right),
\label{EqB}
\end{eqnarray}
where $\delta\left(X {\overset{\alpha}{\rightarrow}}Y\right)$, respectively $\delta\left(X {\overset{\beta}{\rightarrow}}Y\right)$, are equal to $1$ if the transition is induced by the injection of a particle on the left and the ejection of a particle on the right. Using the relations  (\ref{EqCom2}) and (\ref{EqCom3}), one simply obtains  $L^*=R^*=\moyenne{j^*}$, such that
\begin{eqnarray}\label{K-ASEP}
K^*=(L+1)\moyenne{j^*}.
\end{eqnarray}
For $p=q$  and for any set of parameters $\{\alpha,\beta,\delta,\gamma\}$ such that the stationary current vanishes, the system appears to be at equilibrium, characterized by the 'detailed balance' criterion. This situation is obtained when the density of the two reservoirs $\rho_a=\alpha/(\alpha+\gamma)$ and $\rho_b=\delta/(\delta+\beta)$ are equal ($\rho_a=\rho_b=\rho$). In this situation, the steady state is described by the equilibrium (Gibbs) state of the lattice gaz at the density $\rho$ \cite{Derrida-2007}. If the implication $K^*=0\Rightarrow\moyenne{j^*}=0$ was obvious the latter relation states the equivalence $K^*=0\Leftrightarrow\moyenne{j^*}=0$.
\subsection{Open-ZRP}
The zero-range process is a one-dimensional lattice model for which each site may be occupied by an arbitrary number $m$ of particles. In the bulk, particles jump to the next nearest neighboring sites with rates $\theta_m$ function of the occupation number of the departure site. First introduced by Spitzer \cite{Spitzer-1970}, this model has regained interest since the observation of a condensation transition analogous to the Bose-Einstein condensation \cite{Evans-1996}. See \cite{Evans-2000,Evans-2005,Levin-2005,Harris-2005} for a detailed study of the model. We are considering the open-ZRP where particles are added and removed through the boundaries. We note $p\theta_n$ and $q\theta_n$ the left and right transition rates associated to the move of a particle, with $p>q$ such that $\moyenne{j^*}>0$. On the first and last lattice sites, the injection rates of a particle from the reservoirs are respectively given by $\alpha$ and $\delta$.  The ejection of a particle in the reservoirs, occur with rates $q\theta_n$ on the left boundary and $p\theta_n$ on the right boundary. The probability $P^*_X$ of the state $X$, defined by $\{\eta_1(X),\hdots,\eta_L(X)\}$, is factorized under the expression
\begin{eqnarray}
P^*_X=\prod_{n=1}^Lg_n(\eta_n(X)),
\end{eqnarray}
with
\begin{eqnarray}
g_n(m)=\frac{z_n^m}{Z_n}\prod_{k=1}^m\theta_k^{-1},
\end{eqnarray}
where $Z_n$ is the analogue of the grand-canonical partition function
\begin{eqnarray}
Z_n=\sum_mz_n^m\prod_{k=1}^m\theta_k^{-1}.
\end{eqnarray}
It was shown in \cite{Levin-2005} that the fugacities $z_n$ have to satisfy the equality
\begin{eqnarray}\label{EqZRP}
pz_n-qz_{n+1}=\alpha-\gamma z_1=\beta z_L-\delta=\moyenne{j^*}.
\end{eqnarray}
As done previously, we separate the bulk contribution ($B^*$) to the boundary terms ($L^*$ and $R^*$). Considering that the transition $X\rightarrow Y$ consists of the jump forward of a particle initially localized in $n$, we note $m=\eta_n(X)$ and $m'=\eta_{n+1}(X)$. The probabilities $P^*_X$ and  $P^*_Y$ take the form
\begin{eqnarray}
P^*_X&=&g_n(m)g_{n+1}(m') \times\prod_{k\ne n,n+1}g_k(\eta_k(X))\\
P^*_Y&=&g_n(m-1)g_{n+1}(m'+1)\times\prod_{k\ne n,n+1}g_k(\eta_k(X)).\nonumber\\
\end{eqnarray}
In the expression of $B^*$ one has to evaluate the difference
\begin{eqnarray}
p\theta_{m}g_n(m)g_{n+1}(m')-q\theta_{m'}g_n(m-1)g_{n+1}(m'+1).
\end{eqnarray}
With the help of the relation ($\ref{EqZRP}$) we derive the equality $p\theta_mg_n(m)=(\moyenne{j^*}+qz_{n+1})g_n(m-1)$ and $q\theta_{m'+1}g_{n+1}(m'+1)=-(\moyenne{j^*}-pz_{n}-)g_{n+1}(m')$. This finally leads to $B^*=(L-1)\moyenne{j^*}$ and, along the same lines, one show that the contribution of the two boundaries are both exactly equal to the current $L^*=R^*=\moyenne{j^*}$. As for the open-ASEP, the measure of the 'distance' from equilibrium appears to be given by expression ($\ref{K-ASEP}$).\\
One should mentioned that the periodic ZRP can be mapped to a periodic $1D$ asymmetric exclusion process \cite{Evans-2000,Evans-2005}. However, the new process under consideration is not as simple as the ASEP but defined by transition rates depending of the number of empty sites between particles and interpreted as long range interactions. For the periodic ZRP and ASEP one can prove $K^*=L\moyenne{j^*}$. Finally the factor $L$, for periodic systems, or $L+1$, for open systems, is simply the number of possible transitions (between bonds and in/out of the system).\\

Intuitively, for such one-dimensional systems, that do not present other reactions than the biased diffusion, we would conclude that a measure of the 'distance' from equilibrium and violation of 'detailed balance' is given by the calculation of the average macroscopic current. However, a general derivation to any one-dimensional model remain a challenging problem since most of the time the stationary probabilities are unknown.

\section{Outlook and Discussion}
Using the set $\{P^*,K^*\}$ to classify the NESS, we propose a definition of the violation of 'detailed balance' based on the $p$-norm of the matrix formed by the probability currents. If, for finite system sizes, all norms are equivalent, we show by taking the thermodynamic limit of the periodic-TASEP that only $\normK_1$ appears to be a satisfying definition of the 'distance'  from equilibrium ($K^*$). Contrary to the entropy production, this quantity is invariant under the transformations proposed in \cite{RoyceA,RoyceB} that leave the NESS unchanged. In addition, we show that an upper bound of the 'distance' is given by the average of the observable $\AA$ which measure how 'nervous'/'active' the system is. Moreover, we show that both quantities are related to the probabilities of a positive ($\PP_\tau(\delta S>0)$) and negative ($\PP_\tau(\delta S<0)$) creation of entropy $\delta S$ measured over a time interval $\tau<<1$. This results lead explicitly to the expression of $\moyenne{\AA^*}$ and $K^*$ as a function of $\PP_\tau(\delta S=0)$ and $\moyenne{sign(\delta S)}$. For the open-ASEP and open-ZRP, the particular structure of the stationary state allows an analytical calculation which leads to the exact expression $K^*=(L+1)\moyenne{j^*}$. Along the same line, more general results on the violation of 'detailed balance' should be obtained in the near future, applying the matrix product technique to other systems that include charged particles or impurities.\\
Even though the case-by-case analysis reveals $K^*$ as an extensive quantity, at this time, there is no general derivation of this result. We want to motivate the analysis of the 'distance' $K^*$, 'activity' $\moyenne{A^*}$ together with the probability of entropy production $P_t(\delta S)$ measured over an arbitrary time scale. We hope that this study will lead to a better understanding of the measure of the violation of 'detailed balance' and its direct relation with the entropy production.

I would like to thank the group of statistical mechanics at Virginia Tech for its support, especially Professor Zia, Eubank and Kulkarni. I am thankful to D. Karevski, R.J. Harris and C. Maes for helpful discussions, also S.L.Wright, J. Cook, S. Mukherjee and S. Dorosz for their constant encouragement. This research is funded by the US National Science Foundation through DMR-0705152 and the NIH MIDAS project 2U01GM070694-7.


\end{document}